# Dielectric Property of MoS$_2$ Crystal in Terahertz and Visible Region

Xianding Yan,[1] Lipeng Zhu,[1] Yixuan Zhou,[1] Yiwen E,[2] Li Wang,[2] and Xinlong Xu[1,*]

[1]State Key Lab Incubation Base of Photoelectric Technology and Functional Materials, International Collaborative Center on Photoelectric Technology and NanoFunctional Materials, Institute of Photonics & Photon-Technology, Northwest University, Xi'an 710069, China

[2]Beijing National Laboratory for Condensed Matter Physics, and Institute of Physics, Chinese Academy of Sciences, Beijing 100190, China

*Corresponding author: xlxuphy@nwu.edu.cn

Two-dimensional materials such as MoS$_2$ have attracted much attention in recent years due to their fascinating optoelectronic properties. Dielectric property of MoS$_2$ is desired for the optoelectronic application. In this paper, terahertz (THz) time-domain spectroscopy and ellipsometry technology are employed to investigate the dielectric response of MoS$_2$ crystal in THz and visible region. The real and imaginary parts of the complex dielectric constant of MoS$_2$ crystal are found to follow a Drude model in THz region, which is due to the intrinsic carrier absorption. In visible region, the general trend of the complex dielectric constant is found to be described with a Lorentz model, while two remarkable peaks are observed at 1.85 and 2.03 eV, which have been attributed to the splitting arising from the combined effect of interlayer coupling and spin-orbit coupling. This work affords the fundamental dielectric data for the future optoelectronic applications with MoS$_2$.

http://dx.doi.org/10.1364/AO.99.099999

## 1. Introduction

Analogue to graphene, new burgeoning two-dimensional molybdenum disulfide (MoS$_2$) have attracted much attention due to the unique physical, electrical, and optical properties based on the particular energy band structure[1, 2]. MoS$_2$ crystal, which has stacked layer-by-layer structure with inter-plane van der Waals forces, can be considered as the parent of the monolayer MoS$_2$[3-5]. The MoS$_2$ crystal is an indirect band-gap semiconductor with the bandgap energy E$_g$=1.29 eV, while it transformed into a direct semiconductor in the monolayer limit[5]. As a consequence, the electronic property of the MoS$_2$ is more favorable than that of graphene for optoelectronic devices as the lack of bandgap in graphene resulting in low on/off ratios for optoelectronic application with graphene[2]. As such, MoS$_2$ plays an important role in the next generation of optoelectronic devices such as transistors[6], detectors[7], and so on[2] due to the presence of a finite band gap[8].

The optoelectronic applications of materials are strongly determined by the dielectric constant (permittivity), which is the fundamental properties used to characterize refractive index, absorption, conductivity, capacitance, and so on[9]. For MoS$_2$, first-principle calculations have been used to study the optical spectra so far[10, 11], but the juiceful theoretical results call further experimental verification, which is lacking particularly in the terahertz (THz) region. As the working frequency of many optoelectronic and photonic devices climb from GHz to THz range, the THz dielectric response study become more and more important[12, 13], especially for the THz device design with MoS$_2$[14]. Even in the visible region, the experimental data for the dielectric constant of MoS$_2$ is limited. Early works back to 1970s, Liang et al.[15] and Beal et al.[16] presented a reflectivity spectra of MoS$_2$ and deduced the dielectric constant by Kramers-Kronig analysis. Recently, Castellanos-Gomez et al. studied the refractive index of thin MoS$_2$ crystal with the Fresnel law[17]. Liu et al.[18] demonstrated the dielectric constant deduced from the absorption spectrum. However, all these methods are not the direct measurement of the dielectric constant. Spectroscopic ellipsometry, which has been developed into an effective method to measure the dielectric constant directly in a wide research areas from semiconductors to organic materials[19], can be expected to give a clear dielectric property of the MoS$_2$ crystal in visible region.

In this paper, we report the dielectric response of MoS$_2$ crystal in THz and visible region. The dielectric parameters have been extracted by THz time-domain spectroscopy (THz-TDS) and ellipsometry, respectively. In THz region, the real and imaginary parts of the complex dielectric constant of MoS$_2$ can be described with a Drude model, which is due to the intrinsic carrier absorption. In visible region, the general trend of the complex dielectric constant can be described with a Lorentz model. There are two remarkable peaks located at 1.85 and 2.03 eV in dielectric response, which is due to the splitting arising from the combined effect of interlayer coupling and spin-orbit coupling.

This work provides the fundamental data for the future optoelectronic applications with MoS$_2$.

## 2. Experimental section

MoS$_2$ crystal is purchased from SPI Supplies. The thickness of the MoS$_2$ sample is 90 $\mu m$. We selected the MoS$_2$ crystal with a smooth and uniform area for both THz-TDS and spectroscopic ellipsometry in our experiment. A mode-locked Ti:sapphire laser with a center wavelength of 800 nm and a pulse duration of 100 fs was equipped for a home-built THz-TDS[20]. In THz-TDS measurement, we obtain a time-domain THz wave, while the amplitude and phase of the THz wave in frequency domain can be obtained by Fourier transformation. The dielectric property in visible region was measured with a spectroscopic ellipsometer (ESS03, ELLiTOP, Beijing) with multi-wavelength scanning for the ellipsometric parameters (Ψ, Δ), at the incident angle of 70.09°.

## 3. Result and discussion

Figure 1 presents THz time-domain wave of the reference signal (black line) and the sample signal (black line). It is clear that the shape and amplitude of THz signal have been changed as it passes through the sample. The difference between the reference and sample signal indicates that there is strong dielectric response due to the interaction of THz wave with MoS$_2$[21, 22].

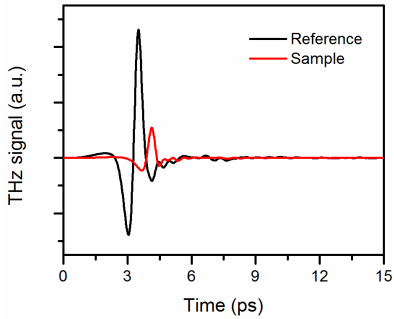

Figure 1. Transmitted THz time-domain signal. The black line for reference signal and the red line for sample signal.

To simplify the parameter extraction procedure, Fabry-Pérot etalon effects have been eliminated by windowing the THz time-domain waves[14]. As such, the complex transfer function for the THz wave through the MoS$_2$ crystal can be expressed as[23]:

$$H(\omega) = \frac{4\tilde{n}(\omega)}{[1+\tilde{n}(\omega)]^2} \exp\{-j\omega[\frac{[\tilde{n}(\omega)-1]d}{c}]\} = A(\omega)e^{-j\phi(\omega)} \quad (1)$$

Where $\tilde{n}(\omega)$ is the complex refractive index, $d$ is the thickness of MoS$_2$ crystal, $c$ is the speed of light in vacuum, $A(\omega)$ is the amplitude of the complex transfer function, and $\phi(\omega)$ is the phase of the complex transfer function. Besides, the complex refractive index can be written as $\tilde{n}(\omega) = n(\omega) + j\kappa(\omega)$, ($n(\omega)$ real refractive index and $\kappa(\omega)$ extinction coefficient). We can obtain the $H(\omega)$ from the reference and sample signals after fast Fourier transformation (FFT). In general, equation (1) has no analytical solution. Newton iterative method is employed, in which the initial value of the refractive index $n(\omega)$ and extinction coefficient $\kappa(\omega)$ are worked out as follows[24]:

$$n(\omega) = \frac{\phi(\omega)c}{\omega d} + 1 \quad (2)$$

$$\kappa(\omega) = \frac{-\ln(A(\omega)[\frac{n(\omega)+1]^2}{4n(\omega)}])c}{\omega d} \quad (3)$$

The relationship between absorption coefficient $\alpha(\omega)$ and the extinction coefficient $\kappa(\omega)$ can be expressed as:

$$\alpha(\omega) = 2\omega\kappa(\omega)/c \quad (4)$$

The calculated absorption coefficient $\alpha(\omega)$ and refraction index $n(\omega)$ are presented in Figure 2. It is evident that the absorption coefficient decreases with the increasing of frequency in the region of 0.5 to 2.0 THz, while it increases a little in the region of 2.0 to 2.3 THz. The refractive index increases with the increasing of frequency in the region of 0.5 to 2.3 THz.

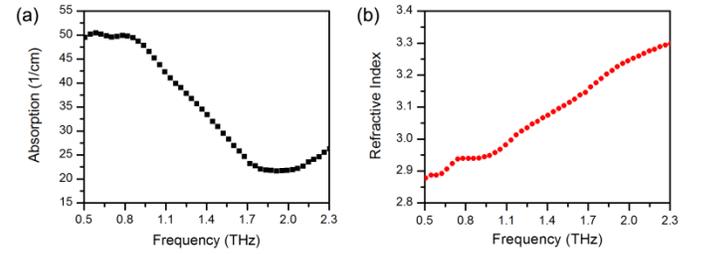

Figure 2. (a) Absorption and (b) refraction index of MoS$_2$ crystal in THz region.

From the absorption and refractive index, the complex dielectric constant $\tilde{\varepsilon}(\omega) = \varepsilon_{real} + i\varepsilon_{imaginary}$ ( $\varepsilon_{real}$, $\varepsilon_{imaginary}$ are the real and imaginary parts of the dielectric constant $\tilde{\varepsilon}(\omega)$, respectively) can be calculated as they have a relationship with the complex refractive index $\tilde{n}(\omega)$ as:

$\varepsilon_{real} = [n(\omega)]^2 - [\kappa(\omega)]^2$ and $\varepsilon_{imaginary} = 2n(\omega)\kappa(\omega)$. As shown in Figure 3, the real part of the dielectric constant increases with the increasing of frequency, while the imaginary part decreases with the increasing of frequency.

The Lorentz model was used to fit the experimental complex dielectric constants for the MoS$_2$ crystal in order to understand the physical mechanism of the dielectric response in THz region[13]. The dielectric function $\tilde{\varepsilon}(\omega)$ in the format of Lorentz response can be described as follows.

$$\tilde{\varepsilon}(\omega) = [n(\omega) - i\kappa(\omega)]^2 = \varepsilon_\infty + \frac{\omega_p^2}{\omega_0^2 - \omega^2 - i\Gamma\omega} \quad (5)$$

$\omega_\infty$ is the high-frequency dielectric constant, and $\omega_p$, $\omega_0$ and $\Gamma$ denoting the oscillator strength, resonant frequency, and damping constant of the mode, respectively. The real and imaginary parts of the dielectric function are presented in Figure 3. The dielectric dependence deviates from the phonon response[25], which is the main response in THz region[24] for semiconductors. Another possible response in THz region is from free carriers[26] for many semiconductors. After fitting with Equation (5), we find that $\omega_0 \approx 0$, which suggests Drude response (Lorentz model reduces to Drude mode as $\omega_0 \approx 0$) from free carriers. This is consistent with the results in Figure 2a, because there is a significant absorption which reduces as the frequency increased. This strongly frequency-dependent absorption is mainly due to the absorption from carriers, which is similar with the analysis in germanium crystalline[26]. The fitting value of the damping rate for the carriers is $\Gamma/2\pi = 1.17$ THz, which is $\Gamma = e/m^*\mu$. Considering the value of carriers mobility is 410 cm²V⁻¹s⁻¹ in MoS$_2$ [2], the calculated $m^* = 0.58 m_0$, which is the averaged electrons and holes. The value is similar to the calculated value of heavier electron effective mass $m^* = 0.45 m_0$ [27] and $m^* = 0.60 m_0$ [28] from ab initio calculations. The fitting value of the plasma frequency $\omega_p/2\pi = 2.67$ THz, which is $\omega_p = \sqrt{Ne^2/\varepsilon_0 m^*}$. And the calculated intrinsic carrier density is $N = 3\times 10^{15} \, cm^{-3}$. The theoretical results are shown as the solid lines in Figure 3, which demonstrates the importance of the intrinsic carriers to the dielectric response of MoS$_2$ in THz region.

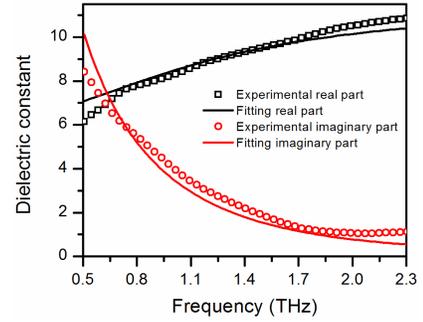

Figure 3. Real and imaginary part of the complex dielectric constant calculated from experimental data (squared dots) as well as the fitting with the Lorentz model (solid lines).

MoS$_2$ crystal is a semiconductor with an indirect band-gap $E_g$=1.29 eV, which can be used as transistors and photodectors in visible region[2]. It is important to characterize the dielectric constant in the visible region with a non-destructive method for the optoelectronic applications. Spectroscopic ellipsometry was applied to measure the extinction coefficient $\kappa$ and refractive index $n$ calculated from the ellipsometric parameters $\Delta$ and $\Psi$. $\Delta$ and $\Psi$ are measured at an incident angle $\varphi$ of 70.09°, which are shown in Figure 4. The wavelength varies from 400 nm to 1600 nm. Except some transition peaks below 750 nm, the $\Psi$ decreases with the increasing of the wavelength, while the $\Delta$ is in the opposite direction. The real and imaginary parts of the dielectric constant of MoS$_2$ can be obtained using the ellipsometry parameters $\Delta$ and $\Psi$ as follows[29]:

$$\varepsilon_{real} = (n_0 \sin\varphi)^2 [1 + \frac{\tan^2\varphi(\cos^2 2\Psi - \sin^2 2\Psi \sin^2\Delta)}{(1+\sin 2\Psi \cos\Delta)^2}] \quad (6)$$

$$\varepsilon_{imaginary} = -\frac{(n_0 \sin\varphi \tan\varphi)^2 \sin 4\Psi \sin\Delta}{(1+\sin 2\Psi \cos\Delta)^2} \quad (7)$$

Where $n_0$ is the refractive index of the environment (air $n_0 = 1.0$) and $\varphi$ is the incident angle of the polarized light.

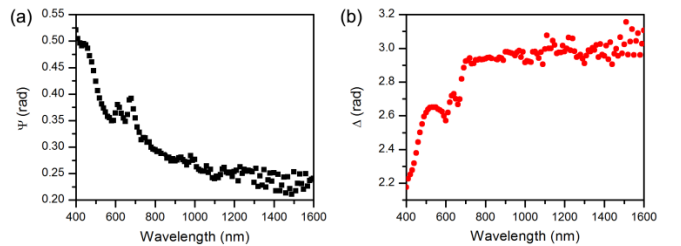

Figure 4. $\Psi$ (a) and $\Delta$ (b) measured by spectroscopic ellipsometry.

The real and imaginary part of the complex dielectric constant by equation (6) and (7) are shown in Figure 5. The real part of the permittivity increases with the increasing of energy until it reaches approximate 1.8 eV and then it decreases with the energy in general. However, the imaginary part of the permittivity increases with the increasing of energy in the range of 0.8 to 2.6 eV and decreases with the energy after 2.6 eV in general. It is also obvious that there are three peaks predominate in the dielectric spectra. As the general tread of both real and imaginary part of the dielectric response follows a Lorentz response, we use Equation 5 to fit both the real and imaginary parts of the complex dielectric constant simultaneously, which is shown as the solid lines in Figure 5. The fitting suggests that the Lorentz model can capture the main background of the real and imaginary parts of the complex dielectric constant.

The Lorentz model parameters are obtained as $\varepsilon_\infty = 4.2$, $\omega_p / 2\pi = 10.06$ eV, $\omega_0 / 2\pi = 2.70$ eV, $\Gamma / 2\pi = 1.35$ eV. As we can see from Figure 5, there is a peak near 2.70 eV, which suggests that the Lorentz model mainly capture the response from $\omega_0 / 2\pi = 2.70$ eV, with the strength as $S = (\omega_p / \omega_0)^2 = 13.9$. There are two additional peaks near 1.85 eV and 2.03 eV, which are mainly due to the leading edge of the strong absorption range. These are interpreted as a spin-orbit split valence band and the splitting value of 0.18 eV are consistent with the previous reports[30].

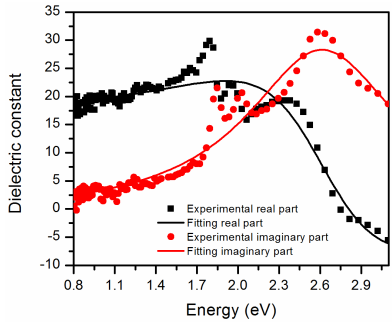

Figure 5. Real and imaginary part of complex dielectric constant of MoS$_2$ crystal in visible region. The squared dots and solid lines are the experiment results and fitting results, respectively.

The absorption coefficient $\alpha$ and refractive index $n$ can be calculated by $n = \sqrt{\dfrac{\sqrt{\varepsilon_{real}^2 + \varepsilon_{imaginary}^2} + \varepsilon_{real}}{2}}$, $\kappa = \sqrt{\dfrac{\sqrt{\varepsilon_{real}^2 + \varepsilon_{imaginary}^2} - \varepsilon_{real}}{2}}$ as shown in Figure 6. It is obvious that the absorption of MoS$_2$ crystal in visible region increases with the increasing of the energy in general. However, the refractive index demonstrates large dispersion in the range of 0.8 to 3.1 eV.

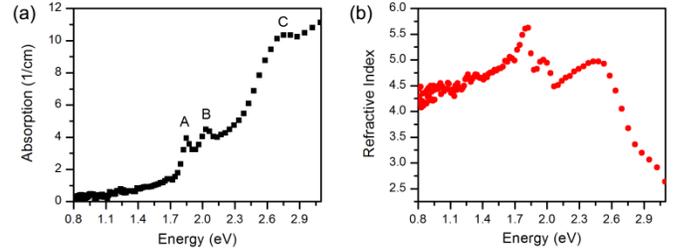

Figure 6. (a) Absorption of MoS$_2$ crystal. (b) Refraction index of MoS$_2$ crystal.

There are two absorption peaks, one locates at 1.85 eV (labeled A in Figure 6(a)) and the other locates at 2.03 eV (labeled B in Figure 6(a)). MoS$_2$ are expected to have strong excitonic effects and Qiu et al. presented first-principles calculations of the optical response of MoS$_2$ and demonstrates excitonic states dominated absorption peaks A and B are located at 1.88 and 2.02 eV[10]. These values are also consistent with the experimental results by Mak et al. with the peaks A and B at 1.88 and 2.03 eV[5]. These two peaks are due to the splitting arising from the combined effect of interlayer coupling and spin-orbit coupling[5]. The transition for peaks A and B are from $\Gamma_9^-$ to $\Gamma_8^+$, and from $\Gamma_7^-$ to $\Gamma_9^+$, respectively[30]. The line-width for peaks A and B are large as the strong coulomb interaction and strong electron-phonon interaction could happen in MoS$_2$. In addition to the observed peaks A and B, there arises another absorption peak located at approximate 2.70 eV, which is consistent with the fitting results in Figure 5. This peak is tentatively attributed to the impurity level related transition as proposed by Evans[31] et al. However, Beal et al. suggested the transitions from high density of states regions at Q point of the two dimensional

Brillouin zone[32]. As this peak happen not only in the experimental data[5, 31-33], but also in the calculated results[10, 28] without considering impurity level, we attribute it to the band transition at Q of the two dimensional Brillouin zone.

## 4. Conclusion

In summary, THz-TDS and ellipsometry technology were used to investigate the dielectric response of $MoS_2$ crystal in THz and visible region. We observed that the absorption of $MoS_2$ crystal decreased with the frequency increasing in THz region, while the refractive index increased with the frequency increasing in THz region. This is mainly due to the response from free carriers, which can be fitted by a Drude model for the complex dielectric constant of $MoS_2$ in THz region. In the visible region, we observed three absorption peaks located at 1.88, 2.02, and 2.7 eV. The general trend of the complex dielectric constant can be described with a Lorentz model due to the absorption peaks at 2.7 eV. The first two peaks are due to the splitting arising from the combined effect of interlayer coupling and spin-orbit coupling, while the third peak are from the band transition at Q of the two dimensional Brillouin zone. The dielectric response of the $MoS_2$ crystal in THz and visible region are useful for the development of $MoS_2$-based optoelectronic devices for THz and visible applications.

## 5. Acknowledgment

This work was supported by National Science Foundation of China (No. 11374240, 11374358), Natural Science basic Research Plan in Shannxi Province of China (No. 2012KJXX-27), Ph.D. Programs Foundation of Ministry of Education of China (No. 20136101110007), Key Laboratory Science Research Plan of Shannxi Education Department (13JS101), National Key Basic Research Program (2014CB339800).